\documentclass[sigconf]{acmart}

\AtBeginDocument{%
  \providecommand\BibTeX{{%
    \normalfont B\kern-0.5em{\scshape i\kern-0.25em b}\kern-0.8em\TeX}}}

\usepackage{pifont}
\usepackage{booktabs} 
\usepackage{amsmath}
\usepackage{multirow}
\usepackage{flushend}
\usepackage{enumitem}
\usepackage[T1]{fontenc}
\usepackage[utf8]{inputenc}
\usepackage{algorithmic}
\usepackage[linesnumbered,ruled,vlined]{algorithm2e}
\usepackage[switch]{lineno}
\usepackage{eqparbox}
\usepackage[nopar]{lipsum}
\usepackage{makecell}

\usepackage{xcolor}

\newtheorem{example}{Example}[section]
\newtheorem*{proof*}{Proof}
\newtheorem*{claim*}{}

\newcolumntype{C}[1]{>{\centering\let\newline\\\arraybackslash\hspace{0pt}}m{#1}}
\newcommand\ChangeRT[1]{\noalign{\hrule height #1}}

\begin{document}

\title{Uni-Retriever: Towards Learning The Unified Embedding Based Retriever in Bing Sponsored Search} 

\author{
    Jianjin Zhang$^{\textbf{{\tiny\ding{168}}}}$,
    Zheng Liu$^{\textbf{{\tiny\ding{168}}}}$,
    Weihao Han$^{\textbf{{\tiny\ding{168}}}}$, 
    Shitao Xiao$^{\textbf{{\tiny\ding{171}}}}$,  
    Ruicheng Zheng$^{\textbf{{\tiny\ding{168}}}}$, 
    Yingxia Shao$^{\textbf{{\tiny\ding{171}}}}$, Hao Sun$^{\textbf{{\tiny\ding{168}}}}$,
    Hanqing Zhu$^{\textbf{{\tiny\ding{168}}}}$,
    Premkumar Srinivasan$^{\textbf{{\tiny\ding{168}}}}$,
    Denvy Deng$^{\textbf{{\tiny\ding{168}}}}$, Qi Zhang$^{\textbf{{\tiny\ding{168}}}}$, Xing Xie$^{\textbf{{\tiny\ding{168}}}}$\\                       
    {\ding{168}: Microsoft WebXT and Microsoft Research Asia}\\
    {\ding{171}: Beijing University of Posts and Telecommunications} \\ 
}
\email{{jianjzh,zhengliu,weihan,ruzhen,hasun,hanqzhu,prsriniv,dedeng,qizhang,xingx}@microsoft.com, {stxiao,shaoyx}@bupt.edu.cn}

\renewcommand{\shortauthors}{Zhang and Liu, et al.}
\renewcommand{\authors}{Zhang and Liu, et. al.}


\begin{abstract}
Embedding based retrieval (EBR) is a fundamental building block in many web applications. However, EBR in sponsored search is distinguished from other generic scenarios and technically challenging due to the need of serving multiple retrieval purposes: firstly, it has to retrieve \textbf{high-relevance} ads, which may exactly serve user's search intent; secondly, it needs to retrieve \textbf{high-CTR} ads so as to maximize the overall user clicks. In this paper, we present a novel representation learning framework \textbf{Uni-Retriever} developed for Bing Search, which unifies two different training modes \textbf{knowledge distillation} and \textbf{contrastive learning} to realize both required objectives. On one hand, the capability of making high-relevance retrieval is established by distilling knowledge from the ``relevance teacher model''. On the other hand, the capability of making high-CTR retrieval is optimized by learning to discriminate user's clicked ads from the entire corpus. The two training modes are jointly performed as a multi-objective learning process, such that the ads of high relevance and CTR can be favored by the generated embeddings. Besides the learning strategy, we also elaborate our solution for EBR serving pipeline built upon the substantially optimized DiskANN, where massive-scale EBR can be performed with competitive time and memory efficiency, and accomplished in high-quality. We make comprehensive offline and online experiments to evaluate the proposed techniques, whose findings may provide useful insights for the future development of EBR systems. Uni-Retriever has been mainstreamed as the major retrieval path in Bing's production thanks to the notable improvements on the representation and EBR serving quality.  
\end{abstract}

\keywords{Sponsored Search, Embedding Based Retrieval, Knowledge Distillation, Contrastive Learning, Representation Learning} 

\maketitle

\section{Introduction}

Embedding based retrieval (EBR) is highly emphasized in recent years \cite{kim-2014-convolutional,karpukhin2020dense,xiong2021approximate,huang2020embedding,liu2021que2search}. In EBR system, queries and answers are represented as latent vectors. The semantic relationship between query and answer is reflected by their embedding similarity; as such, the desirable answers to a query can be efficiently retrieved on top of approximate nearest neighbour search (ANN) \cite{johnson2019billion}. The EBR system is typically learned via contrastive learning \cite{karpukhin2020dense,xiong2021approximate}: given the labeled data, the representation model is learned to discriminate the ground-truth answers to each query from the entire ads corpus. Thanks to the recent progress of deep neural networks, especially the pretrained language models \cite{devlin2018bert}, the accuracy of EBR has been substantially improved, making it a critical recall path in today's web applications, such as online advertising, search engines, and recommender systems. 

\subsection{EBR in Sponsored Search}
Sponsored Search is the major source of income for search engines and online shopping platforms. Given a query from user, the system is expected to retrieve the ads which are not only closely related to user's search intent, but also associated with high CTR in order to maximize the revenue of platform. As a result, EBR in sponsored search is distinguished from EBR in other generic applications, e.g., web search and question answering, due to the need of satisfying multiple retrieval purposes. 
In this place, a toy example is introduced for better illustration of the above scenario. 

\begin{figure}[t]
\centering
\includegraphics[width=1.0\linewidth]{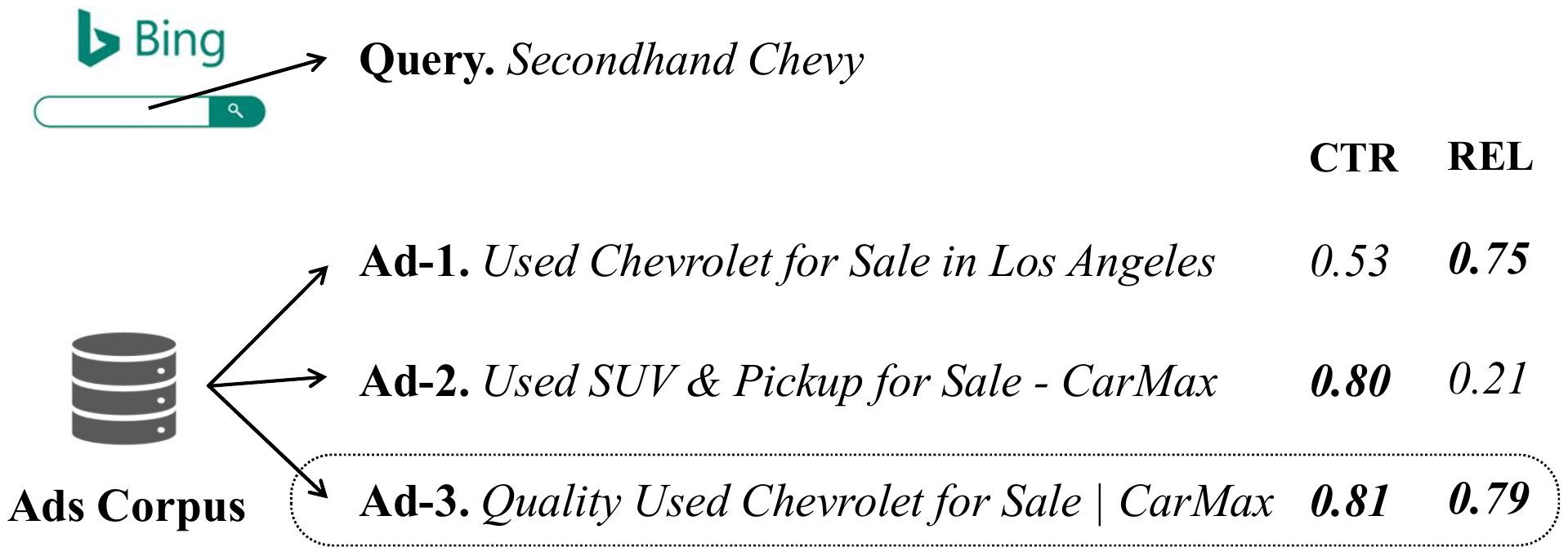}
\caption{Working Example. Ad-3, which is of both high CTR and high relevance with the user's specified query, is expected to be retrieved from ads corpus for sponsored search.}
\label{fig:1}
\end{figure}

\begin{example}\label{example:1}
Suppose the following query is specified by the user: ``{Secondhand Chevy}'' (shown as Figure \ref{fig:1}). Meanwhile, three different ads are presented for sponsored search. \textbf{Ad-1}. ``{Used Chevrolet for sale in Los Angeles}''. Although the first ad is highly relevant with user's query, it suffers from low CTR probably due to the missing of merchant name. \textbf{Ad-2} ``{Used SUV \& Pickup for Sale | CarMax}''. The second ad is associated with high CTR given its popularity to general crowd; however, it is weakly related with the user's specified query. \textbf{Ad-3}. ``{Quality Used Chevrolet for Sale | CarMax}''. The third ad is of both high relevance to the user's query and high CTR thanks to the fascinating quality of the provided service. Therefore, only Ad-3 will be retrieved for the query given that both high-relevance and high-CTR conditions are satisfied by it. 
\end{example} 

The training of the embedding model for sponsored search is non-trivial. Particularly, the typical training algorithms, which mainly rely on contrastive learning, call for a huge collection of queries paired with ground-truth answers. However, such a requirement will be impractical in sponsored search considering that it is almost infeasible to manually label the ``high-relevance \& high-CTR ads'' for a great deal of queries. As a result, most of the previous works tackle this problem with ad-hoc solutions, e.g., the detection of bad cases in \cite{fan2019mobius}. In this work, we develop our unified representation learning framework in Bing Sponsored Search, called \textbf{Uni-Retriever} (the Unified Embedding Based Retriever). The proposed framework jointly leverages two different training modes: \textit{knowledge distillation} and \textit{contrastive learning}, such that the desired embedding model can be effectively trained based on user's click data (which is abundant) and a pretrained relevance teacher model capable of predicting the semantic closeness. 


$\bullet$ \textbf{Knowledge Distillation from Relevance Teacher}. An off-the-shelf teacher model is leveraged for the prediction of the semantic closeness between the query and ad. Although many public sentence representation models, like SBERT \cite{reimers2019sentence} and DSSM \cite{kim-2014-convolutional} may play such a role, we pretrain our in-house relevance teacher model for more precise prediction using a moderate amount of manually labeled query-ad relevance data. While training Uni-Retriever, the teacher’s relevance predictions are required to be preserved by the generated embeddings, i.e., a pair of semantic-close query and ad are expected to result in a large embedding similarity. 

$\bullet$ \textbf{Contrastive Learning with Enhanced Negative Sampling}. Uni-Retriever is learned to retrieve high-CTR ads through contrastive learning. Particularly, it extracts a massive scale of users' queries and ads clicks from the production log, and learns to discriminate the ground-truth to each query (i.e., the clicked ad) from the entire ads corpus. Recent studies indicate that the quality of contrastive learning is largely affected by both scale and hardness of the negative samples \cite{luan2021sparse,qu2021rocketqa}. In Uni-Retriever, we leverage the following two techniques for the enhancement of negative sampling: 1) the Cross-Device negative Sampling (CDS), which significantly augments the scale of negative samples by sharing the cross-device encoding results with gradient compensation; 2) the Relevance-filtered ANN negative Sampling (RAS), which increases the hardness of negative samples by selecting those low-relevance ads from the neighbourhood of query embedding. 

The knowledge distillation and the contrastive learning are linearly combined as a multi-objective learning process, where the generated embeddings are learned to favor both high-relevance ads and user's clicked ads. By doing so, both required conditions in sponsored search can be satisfied by the resulted embedding model. 



\subsection{Serving EBR At Scale}
Aside from the learning of embedding model, another technical challenge, which is more generic in real-world applications, is how to serve the EBR system at scale. Particularly, it is extremely challenging to work with billion-scale embeddings on the computation platforms, meanwhile supporting realtime and high-quality ANN search towards these embeddings. The EBR system is required to trade-off of three factors: 1) time consumption, 2) memory usage, and 3) recall rate. The existing approaches intensively exploits two streams of techniques: 1) the proximity-graph based algorithms, e.g., HNSW, NSG \cite{malkov2018efficient,fu2017fast}, and the vector-quantization based algorithms, e.g., IVFADC, IVFOADC+G+P \cite{jegou2011searching,baranchuk2018revisiting}. The former one is favorable to time efficiency and recall rate, but is likely to result in high memory usage; in contrast, the latter one is memory efficient, but can be relatively limited in time efficiency and recall rate. In this paper, we showcase our scalable and high-quality EBR serving pipeline built upon the substantially enhanced DiskANN \cite{subramanya2019diskann}. 

$\bullet$ \textbf{DiskANN}. We take advantage of DiskANN for its competitive time and memory efficiency. It compresses the ads embeddings via product quantization and leverages Vanama graph (a close variant of NSG) for the quick routing to the approximate nearest neighbours. The storage of the index follows a hierarchical architecture: the tier-1 storage is on RAM, which merely keeps the compressed embeddings; the tier-2 storage is on SSD, which maintains the posting lists recording the dense embeddings and graph connections. The ANN process is divided accordingly: with coarse-grained candidates searched based on the compressed embeddings, and fine-grained result post-verified by the full-precision embeddings.  


$\bullet$ \textbf{Optimization}. The direct usage of DiskANN, though scalable, are prone to severe loss of retrieval quality because of 1) the lossy compression of PQ and 2) the inferior post-verification effect from the original dense embeddings. To mitigate these problems, the following optimization treatments are performed w.r.t. the adopted ANN index. Firstly, the quantization module is jointly learned with the embedding model in order to maximize the retrieval performance resulted from the compressed embeddings. Thanks to the adjusted objective and the end-to-end optimization, the first-stage retrieval quality (i.e., the coarse-grained candidates) can be significantly improved compared with the conventional ad-hoc compression methods, like PQ \cite{jegou2010product} and OPQ \cite{ge2013optimized}. Secondly, the original dense embeddings are further adapted w.r.t. the re-ranking scores for the coarse-grained candidates, such that the top-ranked ads can be better discriminated from the first-stage retrieval result.



The proposed techniques have been integrated as Uni-Retriever step-by-step and mainstreamed into Bing's production for the past period. In this paper, comprehensive offline experiments and online A/B tests are provided for the analysis of the proposed techniques. To the best of our knowledge, this is the first work which elaborates the practice of the latest representation learning algorithms in sponsored search, together with the systematic study of how to best leverage the ANN index for massive-scalable EBR. Corresponding technical discussions and experimental studies
may provide useful insight for both researchers and practitioners in the community. Finally, contributions of this paper are summarized as follows. 
\setlist[itemize]{leftmargin=8.0mm}
\begin{itemize}
    \item We introduce our unified representation learning framework: Uni-Retriever, developed for Bing sponsored search. The proposed framework integrates knowledge distillation and contrastive learning, which enables high-relevance and high-CTR ads to be retrieved for user's query.  
    \item We elaborate our EBR serving pipeline, which inherits the high scalability from DiskANN, and substantially improves the retrieval quality based on the optimization processing. 
    \item We perform comprehensive offline and online experiments for the analysis of Uni-Retriever, whose results verify the effectiveness of our proposed techniques. 
\end{itemize}

\begin{figure*}[t]
\centering
\includegraphics[width=0.67\textwidth]{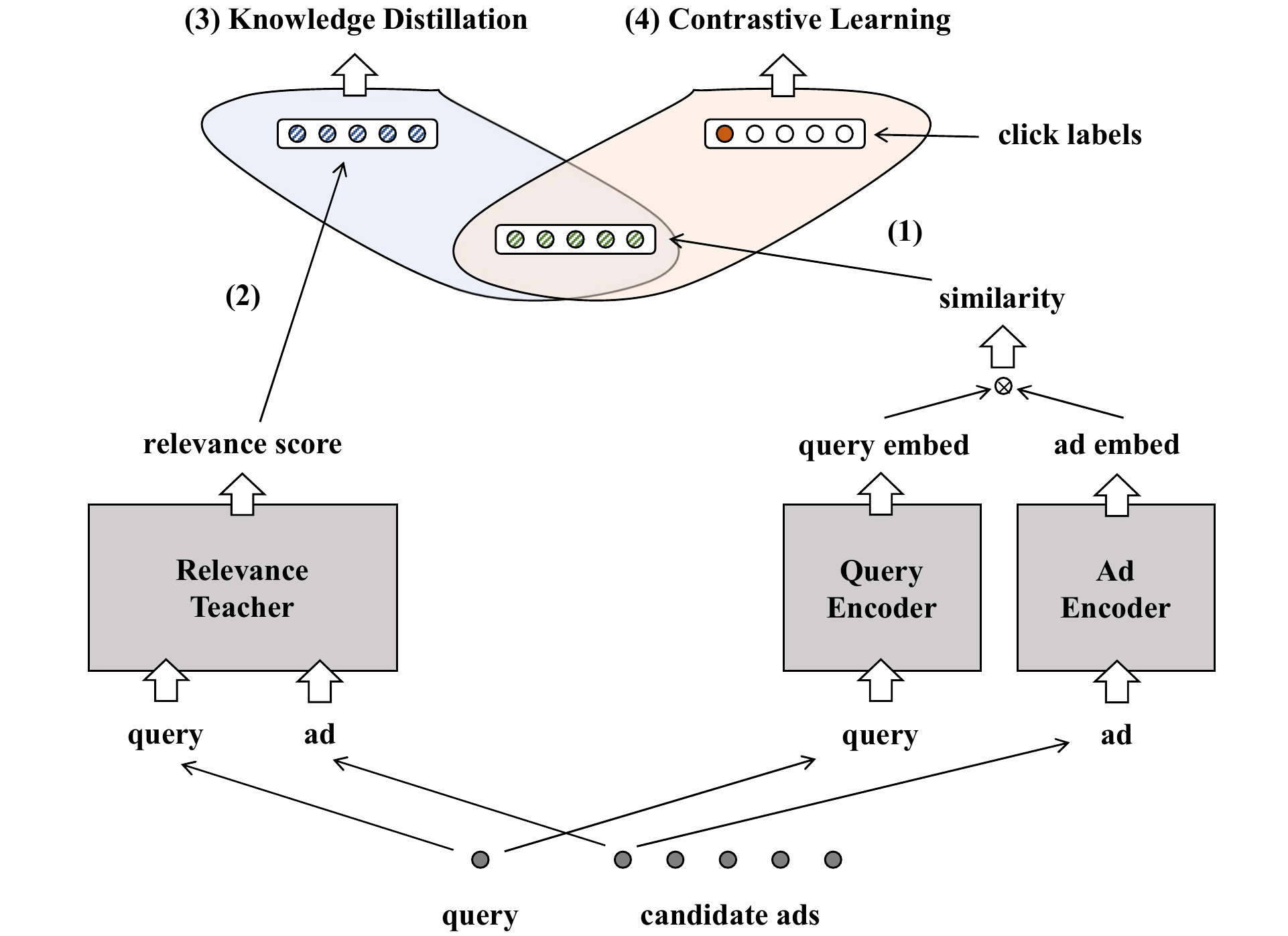}
\caption{Overview of Uni-Retriever. (1) The query and ad are encoded as latent embeddings, whose similarity is computed via inner-product; (2) the semantic closeness (i.e., relevance score) between the query and ad is computed by the relevance teacher model; (3) the knowledge distillation is performed to minimize the difference between the query-ad similarity and the relevance score; (4) the contrastive learning is performed to discriminate the clicked ad.}
\label{frame:1}
\end{figure*} 

\section{Uni-Retriever}
Uni-Retriever is developed to select ads ($A_q$) from the entire corpus ($A$) for user's input query ($q$), which are not only semantically close to the query but also likely to be clicked by the user. The above objective is formulated by the following equation:
\begin{equation}\label{eq:1}
\mathrm{max.} \sum_{A_q} \mathrm{CTR}(q,A_q): ~ s.t. ~
\mathrm{REL}(q,a)\geq\varepsilon, ~\forall a\in{A_q},
\end{equation}
where $\varepsilon$ is the threshold of relevance score. For the ease of optimization, we make the following relaxation to the above objective:
\begin{equation}\label{eq:2}
\mathrm{max.} \sum_{A_q} \mathrm{CTR}(q,A_q) + \lambda*\mathrm{REL}(q,a),
\end{equation}
where $\lambda$ is the positive value trading off the importance between CTR and relevance. In Uni-Retriever, $A_q$ is retrieved based on MIPS (Maximum Inner Product Search). That's to say, the group of ads with the top-$K$ embedding similarities to the query (measured by inner product) are expected to be the optimal solution of Eq. \ref{eq:2}. Such an embedding model is learned by our compound representation learning strategy (shown as Figure \ref{frame:1}), which unifies knowledge distillation and contrastive learning. 

\subsection{Knowledge Distillation}

Firstly, we expect high-relevance ads to be favored by the learned embeddings. To this end, we propose to learn query and ad embedding by distilling knowledge from the relevance teacher model. 

$\bullet$ \textbf{Relevance Teacher Model}. The relevance teacher model is a BERT based binary classification model (denoted as $\mathrm{BERT}^{Tch}$), which predicts whether the given query and ad are semantically close (e.g., label ``0''/``1'' indicate the ad to be non-relevant or relevant with the input query). Unlike the embedding model which adopts the bi-encoder architecture, the relevance teacher model follows the cross-encoder, which is more expressive and capable of making more precise classification. Without loss of generality, the computation of relevance score is given as: 
\begin{equation}\label{eq:3} 
    \text{Rel}_{q,a} = \sigma\big(W^T\mathrm{BERT}^{Tch}([\text{CLS},Query,\text{SEP},Ad])\big),
\end{equation}
where $[\text{CLS},Query,\text{SEP},Ad]$ means the concatenation of query and ad, with CLS and SEP token padded; we take the final layer's hidden state corresponding to CLS as the output of BERT; $W$ ($\in\mathrm{R}^{d\times1}$) is the linear projection which reduces the output vector to a real value; and $\sigma(\cdot)$ is the sigmoid activation. The relevance teacher model is pretrained based on manually labeled data from human experts. 

$\bullet$ \textbf{Knowledge Distillation}. Based on the well-trained relevance teacher model, Uni-Retriever can be learned by knowledge distillation \cite{hinton2015distilling} (in this place, the same backbone model is shared by the query encoder and ad encoder, denoted as $\mathrm{BERT}^{Uni}$). Particularly, the relevance teacher model is used to generate the relevance score $\text{Rel}_{q,a}$ for the input query $q$ and ad $a$. Then, $\mathrm{BERT}^{Uni}$ is learned to imitate the teacher's prediction with the inner product of the query and ad embedding, where the following loss is minimized: 
\begin{equation}\label{eq:4}
    \mathrm{min.}\sum_q\sum_a\big\| \text{Rel}_{q,a} - \langle\mathrm{BERT}^{Uni}(q),\mathrm{BERT}^{Uni}(a)\rangle \big\|. 
\end{equation}
In this place, $\langle\cdot\rangle$ indicates the inner product operator. With the optimization of the above problem, the high-relevance ads to the query can be favored by high inner products, therefore enabling them to be retrieved via MIPS. 

\subsection{Constrastive Learning}

We expect the high-CTR ads to be promoted by the learned embeddings as well. To this end, the contrastive learning is performed, where the embedding model is learned to discriminate the ground-truth (i.e., the clicked ad) for each query from the rest of the ads within the corpus. In this place, the following InfoNCE loss is formulated as our learning objective: 
\begin{equation}\label{eq:5}
    \mathrm{max}.\sum_{q} \frac{\exp(\langle\mathrm{BERT}^{Uni}(q),\mathrm{BERT}^{Uni}(a_q^{+})\rangle)}
    {\sum_{a^- \in N_q}
    \exp(\langle\mathrm{BERT}^{Uni}(q),\mathrm{BERT}^{Uni}(a^{-})\rangle)},
\end{equation}
where $a_q^+$ denotes the ground-truth, and $N_q$ are the negative samples to the query. The contrastive learning's performance is highly affected by two factors: 1) the scale of negative samples \cite{chen2020simple,he2020momentum}, and 2) the hardness of the negative samples \cite{xiong2021approximate,luan2020sparse}. In our project, the following strategies are adopted for the enhancement of scale and hardness of the negative samples.

\begin{figure}[t]
\centering
\includegraphics[width=1.0\linewidth]{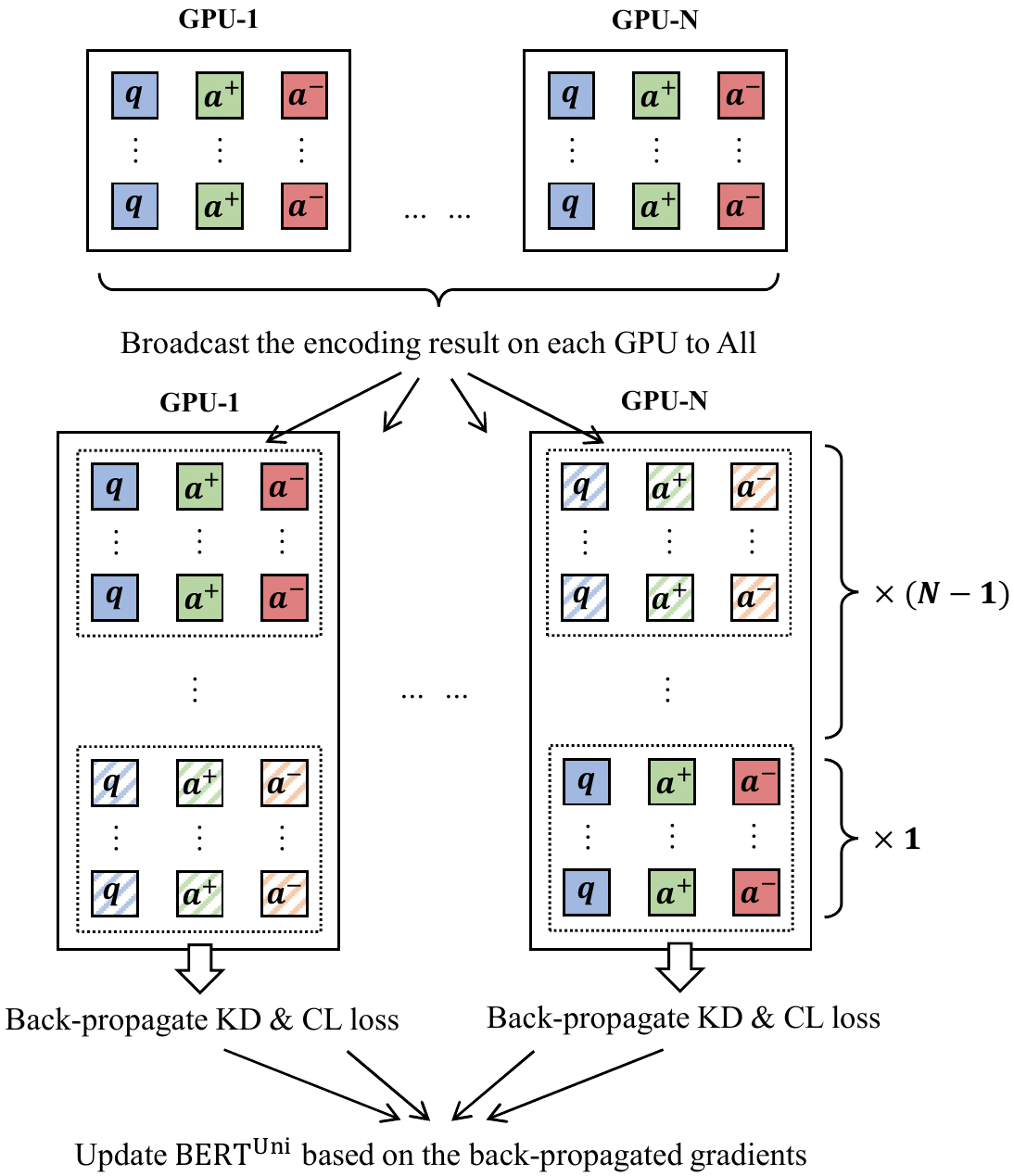}
\caption{Cross-device negative sampling with gradient compensation. (Blue/Green/Red: query embedding/ground-truth embedding/hard-negative embedding. Solid rectangles: locally encoded embeddings, which are differentiable; shadowed rectangles: the broadcasted embeddings from other GPUs, which are non-differentiable).}
\label{fig:2}
\end{figure}

\subsubsection{Cross-device Negative Sampling} The in-batch negative sampling is a widely used strategy which introduces a large number of negative samples in a cost-free manner. Particularly, one query $q$ will use the ground-truth ads of other queries within the same mini-batch $B_q$ as its negative samples: $N_q = \{a^+_{q'\neq{q}}\}_{B_q}$; as a result, each query may get as many as $|B_q|-1$ negative samples. The cross-device negative sampling is a further augmentation of the conventional in-batch negative sampling \cite{xiao2021match,qu2021rocketqa} when the model is trained on multiple distributed GPU devices (where the entire collection of mini-batches is denoted $B = \{B_1,...,B_N\}$): the ground-truth ads from other queries, not only within the same device ($B_q$) but also from the whole devices ($B$), will be used as the negative samples of query $q$; in other words, $N_q = \{a^+_{q'\neq{q}}\}_{B}$. Such an operation will enlarge the negative sample size by $N$-fold without introducing additional computation cost (e.g., $8\times$ larger than in-batch negative sampling when 8 GPUs are used for distributed training). 


It should be noted that the naive cross-batch negative sampling is problematic and will result in limited performance \cite{xiao2021match}; this is because the embeddings from other devices are detached from the computation graph, thus non-differentiable. In this place, we leverage the following \textbf{gradient compensation} operation, with which all cross-device embeddings are make ``virtually differentiable'' so that the model to be correctly optimized (shown as Figure \ref{fig:2}). Firstly, the entire encoding output, i.e., queries embeddings ($\{\mathrm{BERT}^{Uni}(q)\}_B$) and ads embeddings ($\{\mathrm{BERT}^{Uni}(a)\}_B$), will be broadcasted over all the devices ($B$). Secondly, for each of the devices $B_i$, the InfoNCE loss is computed for all the queries (i.e., not only the queries encoded on $B_i$, but also those broadcasted queries from other devices), with all ads embeddings (both locally encoded and globally broadcasted) taken as the negative samples. Finally, the InfoNCE losses are back propagated from all the devices and reduced for the update of the model. It is straightforward that the above processing is equivalent 
By doing so, the partial gradients from the rest of the GPUs will compensate for the non-differentiable ads embeddings on one GPU. Therefore, the InfoNCE losses generated on different devices will correctly update the model parameters where the related queries and ads embeddings are encoded. 


\subsubsection{Relevance Filtered Hard Negatives}. In addition to the cross-device negative samples, hard negatives are also introduced for each query. Compared with other heuristically acquired hard negatives, the \textbf{ANN based hard negatives} are proved to be much more useful in contrastive learning \cite{gillick2019learning,xiong2020approximate,qu2021rocketqa}. Particularly, for each query $q$, the ads within $q$'s neighbourhood are sampled as the hard negatives, $a: \mathrm{BERT}^{Uni}(a) \in \text{ANN}(\mathrm{BERT}^{Uni}(q))$. The conventional ANN hard negatives are randomly sampled from the  $q$'s neighbourhood. In our scenario, we expect high-relevance ads to be promoted; therefore, we propose to \textbf{filter the sampling by relevance score}: the neighbouring ads (e.g., the Top-$200$ nearest neighbours) are firstly ranked by their relevance scores; then, the top ranked ads (e.g., Top-1$\sim$K) are removed; finally, the sampling is performed within the remaining ads (e.g., Top-(K+1)$\sim$200). With this operation, we may select those hard yet low-relevance ads as our negatives, which contributes to the representation quality. We empirically find that sampling no more than 4 ANN hard negatives from the filtered neighbours will be sufficient for the best recall performance; further increasing of the ANN hard negatives leads to extra training cost but shows no more improvement. The hard negatives of one query will also be shared with other queries, which may further augment the scale of negative samples. 

\subsection{Disentangle and Multi-objective Learning}
The embeddings are \textbf{disentangled} w.r.t. the relevance and CTR objective for better empirical performances. Particularly, instead of directly using the output of $\mathrm{BERT}^{uni}$, different pooling heads are used for the two objectives. While optimizing the relevance objective in knowledge distillation, the output embedding becomes $W_\text{rel}\mathrm{BERT}^{uni}(\cdot)$; and for the CTR objective in contrastive learning, the output embedding is $W_\text{ctr}\mathrm{BERT}^{uni}(\cdot)$. (Note that the encoding backbone $\mathrm{BERT}^{uni}$ remain shared for the two objectives.) 

Uni-Retriever is jointly trained with knowledge distillation and contrastive learning. Given a mini-batch of triplets: \{query: $q$, positive ad: $a^+_q$, hard-negative ad: $a^-_q$\} (the algorithm remains the same when working with more than one hard negatives), Uni-Retriever will represent them as their CTR and relevance embeddings, respectively; and compute their inner-product similarities for CTR and relevance. Then, the relevance teacher model will be used to predict the relevance scores between the input queries and ads, based on which the knowledge distillation loss can be computed. Note that the relevance teacher model is an cross-encoder, whose encoding cost will be formidable when all input ads need to processed. Therefore, the knowledge distillation will only be performed w.r.t. the clicked ads for the sake of high cost-efficiency. Meanwhile, the contrastive loss will also be computed, where $N_q$ will be composed of the hard-negative of $q$ and the cross-device negatives. Both losses will be added up and back-propagated to update the shared encoding backbone $\mathrm{BERT}^{Uni}$ and the individual pooling heads: $W_\text{rel}$, $W_\text{ctr}$. Both pooling heads are $d \times d$; and the two output embeddings $W_\text{rel}\mathrm{BERT}^{uni}(\cdot)$ and $W_\text{ctr}\mathrm{BERT}^{uni}(\cdot)$ will be normalized and added up as a single vector for EBR.  

\section{EBR Serving Pipeline} 
In Bing Sponsored Search, Uni-Retriever needs to serve billion-scale of ads from different advertisers. Conventional solutions, such as HNSW, NSG, IVFPQ, will be limited by either the huge memory cost or severe loss of retrieval quality. To confront this challenge, we leverage DiskANN \cite{subramanya2019diskann} as our ANN index, where the proximity graph and product quantization are combined for competitive time and memory efficiency. We further introduce a couple of optimization processing w.r.t. the index: MoPQ \cite{xiao2021match} and Re-ranking oriented adaptation, which substantially improve the retrieval quality for the first-stage retrieval and post-verification. 

\subsection{DiskANN}

The ads embeddings are organized by Vamana graph \cite{subramanya2019diskann}, a close variant of NSG \cite{fu2017fast} with parameterized RNG condition for better connectivity of the graph nodes. Based on such a component, the input query can be routed to its nearest ads with high time efficiency. Given that the ads embeddings and their graph connections are too large to fit into memory, the following hierarchical storage architecture is adopted. 

$\bullet$ \textbf{Tier-One Storage}. The tier-one storage is hosted in RAM. To accommodate the entire corpus, the ads embeddings are compressed based on product quantization (more effective solution to be discussed in the next part). The PQ compressed embeddings can be one or two magnitudes lighter than the original dense embeddings; therefore, billion-scale of ads can be hosted in the main memory with tens of GB RAM usage. 

$\bullet$ \textbf{Tier-Two Storage}. The tier-two storage is hosted in SSD for the posting lists, where each entry keeps the following three information: the ID of each ad (int32/64), the full-precision dense embedding of each ad (d-dim float32 vector), and the list of IDs for the neighbouring ads on Vamana graph (int32/64). 

$\bullet$ \textbf{Two-steps ANN Search}. The ANN search is performed with two consecutive steps: coarse-grained search, and fine-grained post-verification. For coarse-grained search, the input query is routed on Vamana graph to get its approximate nearest neighbours. Starting from the entry point, it continues to explore the close yet unvisited neighbours, and add the nearest candidates to a fix-sized priority queue. The exploration on graph is guided by the distance computed with the compressed embeddings. Once an graph node is visited, its neighbours' IDs will be loaded to RAM for future exploration. The full-precision dense embeddings of the visited nodes will also be loaded into RAM. The post-verification is performed once the exploration finishes, where candidates within the priority queue will be refined based on their dense embeddings (already loaded in RAM during the exploration, thus avoiding extra I/O operations). The Top-K nearest ads will be returned as the final result. 

\begin{figure}[t]
\centering
\includegraphics[width=1.0\linewidth]{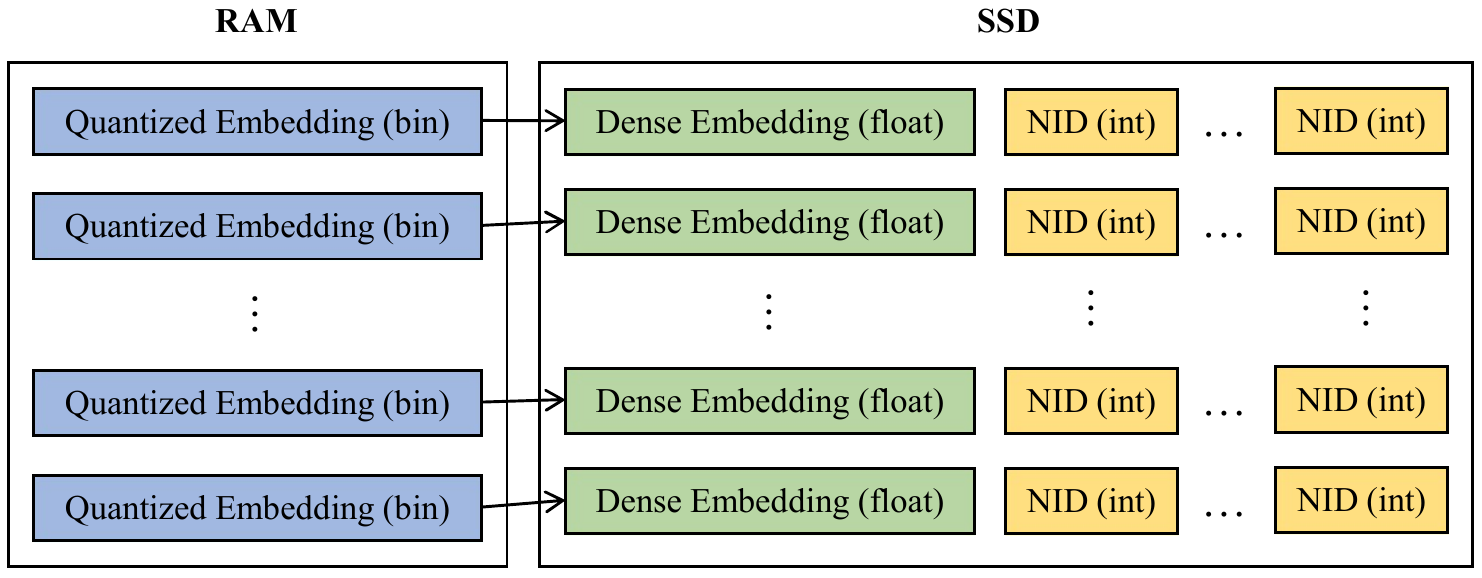}
\caption{Illustration of the two-tier storage. The lightweight PQ compressed embeddings are stored in RAM. The full-precision dense embeddings and the graph connections (NID: Neighbour's ID) are stored in SSD. The compressed embeddings are used for in-memory routing; once a node is visited, the dense embeddings and neighbour IDs will be loaded to RAM for post-verification and the next-round routing.} 
\label{fig:3}
\end{figure}

\begin{figure*}[t]
\centering
\includegraphics[width=0.80\textwidth]{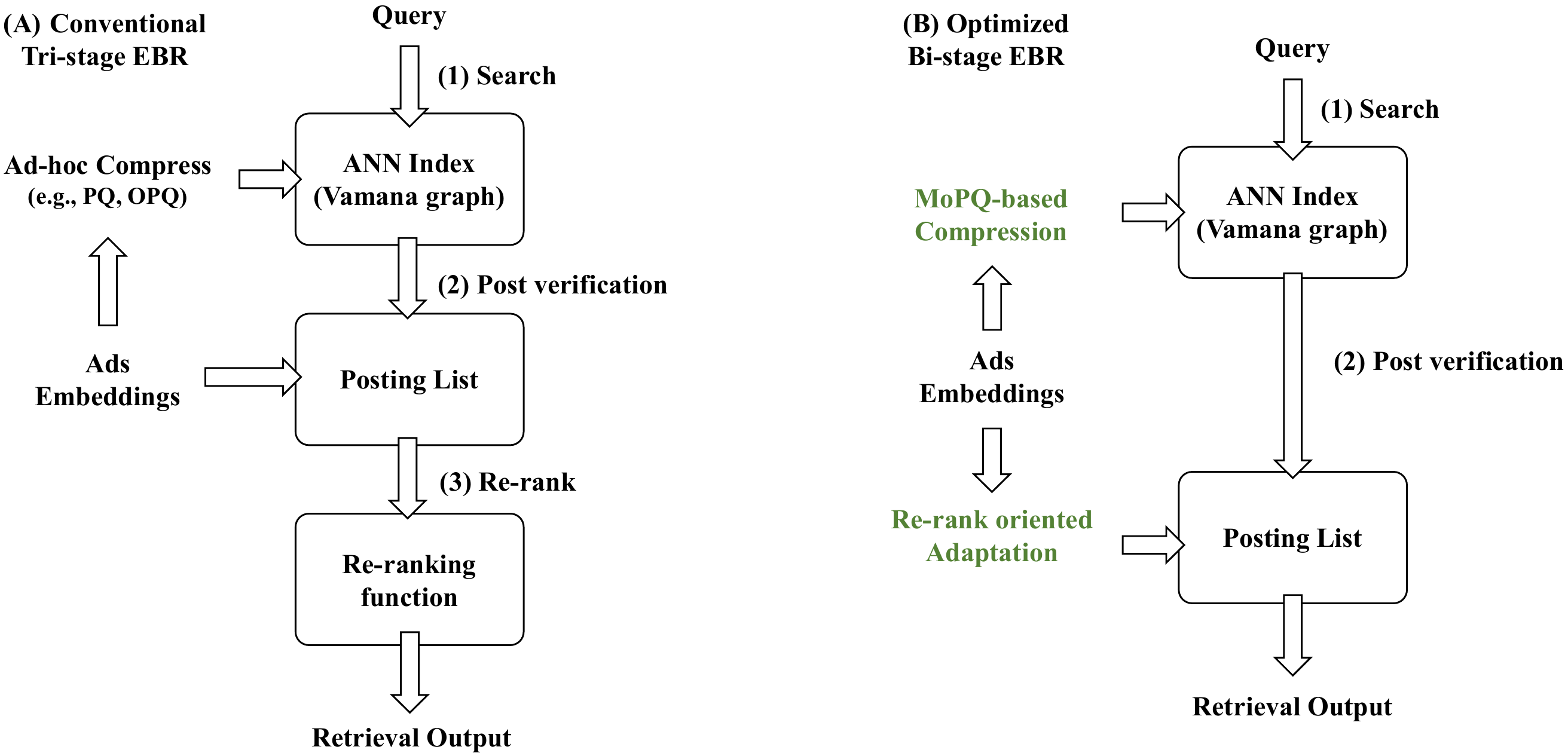}
\vspace{-5pt}
\caption{Comparison between (A) the conventional tri-stage EBR and (B) the optimized bi-stage EBR. The re-ranking operation is omitted in (B) as it is virtually moved ahead to the post-verification step thanks to the re-rank oriented adaptation.} 
\vspace{-5pt}
\label{fig:4}
\end{figure*}

\subsection{Optimization Processing} 

The conventional DiskANN-based EBR goes through the Tri-stage workflow shown as Figure \ref{fig:4} (A). Firstly, the query searches for its coarse-grained candidate ads based on the Vamana graph and the compressed ads embeddings generated by ad-hoc compression methods, e.g., PQ or OPQ. Secondly, the coarse-grained candidates are post verified by the full-precision ads embeddings loaded from SSD to RAM. Finally, the fine-grained candidates are further re-ranked in order to generate the final retrieval output. Notice that the re-ranking function\footnote{The concrete formula of $\Psi$ is kept confidential due to business concern.} is introduced for the optimal business outcome, which combines three factors: CTR and Relevance (Rel) for the query and ad, and Bidding price of the ad (Bid):  
\begin{equation}\label{eq:7}
    \text{Rank}_{q,a} = 
    \Psi(\text{CTR}_{q,a}, \text{Rel}_{q,a}, \text{Bid}_a). 
\end{equation}
In our system, we optimize the conventional tri-stage workflow with the following adaptations: 1) leveraging the \textbf{Matching oriented Product Quantization} (MoPQ) for high-quality first-stage retrieval, and 2) the \textbf{re-ranking oriented adaptation} of the dense embeddings for more effective post verification. 

\subsubsection{MoPQ} The typical quantization methods used by ANN index, e.g., PQ \cite{jegou2010product,jegou2011searching,subramanya2019diskann}, OPQ \cite{ge2013optimized}, ScaNN \cite{guo2020accelerating}, are learned to minimize the reconstruction loss between the original dense embeddings and the PQ compressed embeddings. However, the reduction of reconstruction loss does not necessarily lead to the improvement of retrieval performance (as proved by \cite{xiao2021match}), which will severely limit the retrieval quality under large compression ratios. Our system leverages MoPQ \cite{xiao2021match}, the state-of-the-art product quantization where the compressed embeddings are learned to optimize the retrieval performance: for each ad embedding $a$, the compressed embedding $\hat{a}$ is generated based on codebooks $\mathbf{C}$ ($M \times P$): 
\begin{equation}\label{eq:8}
    \hat{a} = \mathrm{concat}([c_1,...,c_M]): ~
    c_i = \mathrm{argmin}_{j=1...P}(\|a_i - C_i^j \|),
\end{equation}
where $a_i$ is the $i$-th segment of ${a}$. The straight through estimator is used for the codeword selection, which maintains the computation differentiable. The compressed ads embeddings are learned together with the original dense embeddings, following the same objectives as Eq. \ref{eq:4} and \ref{eq:5}. Thanks to the adjusted objective and the collaboration between the embedding model and codebooks, the first-stage retrieval performance can be substantially enhanced by the compressed embeddings. 



\subsubsection{Re-ranking oriented posting list} 
The original dense embeddings are learned to retrieve high-relevance and high-CTR ads from the entire corpus. Therefore, they are inferior for the ranking purpose considering that 1) the re-ranking function is different from the retrieval objective, 2) the re-ranking emphasizes the local discrimination of the candidates from the first-stage retrieval, rather than the global discrimination of the ads within the entire corpus. To mitigate the problem, we further adapt Uni-Retriever for better post-verification by distilling knowledge from the re-ranking function, where the following loss is minimized: 
\begin{equation}\label{eq:9}
    \mathrm{min.}\sum_q\sum_{\hat{A}_q}\big\| \text{Rank}_{q,a} - \langle\mathrm{BERT}^{rank}(q),\mathrm{BERT}^{rank}(a)\rangle \big\|. 
\end{equation} 
In this place, $\mathrm{BERT}^{rank}$ is the BERT model adapted from the original $\mathrm{BERT}^{uni}$, $\hat{A}_q$ are the sampled ads from the first stage retrieval result, $\text{Rank}_{q,a}$ is the ranking score predicted in Eq. \ref{eq:7}. 

\subsubsection{Optimized bi-stage retrieval} With the above optimized processing, EBR is simplified into the bi-stage workflow shown as Figure \ref{fig:4} (B). There are two notable changes compared with the conventional tri-stage workflow. Firstly, the ads embeddings are compressed by MoPQ for the enhancement of the first-stage retrieval. Secondly, the dense embeddings in posting lists are adapted to the re-ranking function; therefore, the re-ranking operation is omitted as it is virtually moved ahead to the post-verification step. 

\begin{table*}[h]
    \centering
    \small
    \begin{tabular}{p{2.4cm} | C{1.6cm} C{1.6cm} C{1.6cm} | C{1.6cm} C{1.6cm} C{1.6cm} }
    \ChangeRT{1pt}  & 
    \multicolumn{3}{c|}{\textbf{Click}} & 
    \multicolumn{3}{c}{\textbf{Relevance}} \\
    \cmidrule(lr){1-1}
    \cmidrule(lr){2-4}
    \cmidrule(lr){5-7}
     \textbf{Method} & 
     \textbf{Hit@10} & \textbf{Hit@100} & \textbf{Hit@200} & \textbf{Rel@10} & \textbf{Rel@100} & \textbf{Rel@200} \\
     \hline
     Baseline & 0.2079 & 0.4092 & 0.5810 & 0.7260 & 0.6971 & 0.6675 \\
     + Multi-obj & 0.2347 & 0.4188 & 0.5841 & 0.7958 & 0.7481 & 0.7348 \\
     + Disentangle & 0.2441 & 0.5122 & 0.5944 & 0.7998 & 0.7553 & 0.7340 \\
     + In-batch & 0.2773 & 0.5263 & 0.6004 & 0.7753 & 0.7191 & 0.6999 \\
     + Cross-device & 0.2887 & 0.5358 & 0.6080 & 0.7833 & 0.7248 & 0.7050 \\
     + ANN negative & 0.3151 & 0.5750 & 0.6455 & 0.8344 & 0.7757 & 0.7527 \\
     + Score filter & 0.3156 & 0.5755 & 0.6464 & \textbf{0.8406} & \textbf{0.7827} & \textbf{0.7596} \\
     + Rank filter & \textbf{0.3286} & \textbf{0.5896} & \textbf{0.6574} & 0.8372 & 0.7755 & 0.7509 \\
    \ChangeRT{1pt}
    \end{tabular}
    \caption{Evaluation of the retrieval quality: the proposed techniques in Uni-Retriever are added step-by-step. 
    The integration of all techniques: ``+ Score filter'' and ``+ Rank filter'' (with the utilization of disentangled embeddings, cross-devices negative sampling, and ANN negatives filtered by relevance-score / relevance ranking-order) achieve the best overall performance.
    }
    \vspace{-20pt}
    \label{tab:1}
\end{table*}

\section{Experimental Studies} 
The offline experiments and online A/B tests are performed to clarify the following problems: 1) the effectiveness of the proposed representation learning framework, 2) the effectiveness of the proposed EBR serving pipeline, 3) the impact to online production. 

\subsection{Experiment Settings}
\subsubsection{Data preparations and Training details} 
A large-scale dataset is curated from Bing's production for offline experiments. There are a total of $1.5$ billion query-ad pairs from real-world users, where $5.7$ million of them are held out for testing. Besides, 20 million ads are included by the ANN index for EBR evaluation. Our text encoder follows BERT base architecture \cite{devlin2018bert}; the output embedding's dimension is linearly projected to 64 from 768. We leverage WordPiece as our tokenizer \cite{wu2016google}, which is trained based on the in-house ads corpus for better empirical performances \cite{yao2021adapt}. There are a total of $50,777$ vocabularies for the well-trained tokenizer. The models are implemented with python 3.6 and pytorch 1.7.0, and trained on machines with 4* NVIDIA-V100-32G GPUs and 2* AMD EPYC 7V12 64-Core CPUs. 


\subsubsection{Evaluation metrics}
The following two metrics are utilized to evaluate retrieval quality: 1) \textbf{Recall of click} (Hit@K): which measures whether the clicked ad can be included by the Top-K retrieval result. 2) \textbf{Relevance degree} (Rel@K): which measures the average relevance score for the Top-K retrieval result. Note that the relevance is predicted by the high-precision relevance teacher, whose accuracy is comparable to human annotator. One more metric is added to evaluate the retrieval quality after re-ranking: 3) \textbf{Re-ranking performance} (NDCG, MRR), which measures the post re-ranking quality for the retrieval result w.r.t. the ground-truth ads selected by the re-ranking function from the entire corpus.

\subsubsection{Method to compare} 
The proposed techniques are added to the Baseline one-by-one for the analysis of their individual impact. 

\setlist[itemize]{leftmargin=5.0mm}
\begin{itemize}
    \item \textbf{Baseline}, which is the most basic approach purely based on contrastive learning (i.e., without knowledge distillation from the relevance teacher model). Besides, each query is sampled with 10 random negatives from the ads corpus. 
    \item \textbf{+ Multi-obj}, which improves the ``Baseline'' by adding the knowledge distillation task. 
    \item \textbf{+ Disentangle}, which improves ``+ Multi-obj'' by having the embeddings disentangled w.r.t. different objectives in the training stage, and added up as one vector for EBR serving. 
    \item \textbf{+ In-batch}, which improves  ``+ Disentangle'' by replacing the random negative samples with the in-batch negative samples.
    \item \textbf{+ Cross-device}, which switches the in-batch negative samples to cross-device negative samples for the above method.
    \item \textbf{+ ANN negative}, which further improves ``+ Cross-device'' by adding one more hard negative sample for each query from ANN (randomly selected from the Top-200 ANN result). 
    \item \textbf{+ Score filter}, which improves ``+ ANN negative'' by using hard yet low-relevance negative samples. Particularly, the ads with relevance scores (normalized within 0 and 1) below threshold (set as 0.5 by experiment) are filtered from the Top-200 ANN result for negative sampling.
    \item \textbf{+ Rank filter}, which also uses hard yet low-relevance negative samples. Different from ``+ Score filter'', the Top-200 ANN result is ranked by the relevance teacher in the first place; then, the lower half (101$\sim$200) is used for negative sampling. 
\end{itemize}

\subsection{Experiment Analysis} 

\subsubsection{Analysis of retrieval quality} The experiment result is shown as Table \ref{tab:1}, where the cumulative impact from our propose techniques are analyzed. The Baseline is based on the simplest contrastive learning method, which leads to the lowest performance. By adding our proposed techniques step-by-step, the overall performances can be gradually improved. Finally, the integration of all proposed techniques, i.e., ``\textbf{+ Score filter}'' and ``\textbf{+ Rank filter}'', which jointly leverage ``\textit{the combination of knowledge distillation and contrastive learning}'', ``\textit{cross-device negative sampling}'', and ``\textit{ANN hard negatives filtered by relevance score / relevance ranking-order}'', achieve the most competitive performances. More detailed findings towards each of the techniques are analyzed as follows.


Firstly, with the combination of knowledge distillation (optimizing the relevance) and contrastive learning (optimizing CTR), both metrics: Click and Relevance, can be substantially improved by ``+ Multi-obj'' compared with the Baseline. This finding indicates 1) the proposed multi-objective learning framework is effective, 2) the optimization of relevance and CTR objectives are not contradicted, which can be mutually reinforced to a large extent. Besides, by generating disentangled embeddings (+ Disentangled) for the two objectives, the representation quality can be further improved against our basic multi-objective learning method. 

Secondly, by using in-batch negatives (+ In-batch), the click metrics (Hit@K) can be significantly improved over the previous methods, which rely on 10 randomly sampled negatives. And with cross-device negative sampling (+ Cross-device), the click metrics can be further improved. Such observations verify that the increased scale of negative samples is beneficial to the representation quality. Note that the relevance metrics are not improved, probably because both methods are simply applied to the contrastive learning. 

Thirdly, the overall performances can be continually improved by introducing the ANN hard negatives (+ ANN negative). Different from the ``+ In-batch'' and ``+ Cross-device'' which augment the scale of negative samples, it is quite interesting that both click and relevance metrics may benefit from the ANN hard negatives. Moreover, additional improvements can be achieved with ``+ Score filter'' and ``+ Rank filter'', as the high-relevance ads can be effectively excluded from being selected as negative samples. 

\subsubsection{Impact from Re-rank oriented Adaptation} The impact from the re-rank oriented adaptation is shown in Table \ref{tab:2}. ``Base'' stands for the conventional post-verification method, where the original dense embeddings are utilized; ``Base + Re-rank'' indicates that the post-verification result is further re-ranked; ``Adapt to Re-rank'' is our method, where the post-verification is made based on dense embeddings adapted to the re-ranking function. It can be observed that ``Adapt to Re-rank'' notably outperforms ``Base''. Such a finding is expected, considering that with adaptation to the re-ranking function, the candidate ads with high re-ranking scores can be further promoted by the dense embeddings during post-verification. Besides, it is interesting that ``Adapt to Re-rank'' also outperforms ``Base + Re-rank''. This is because a great deal of high-quality candidates are left out due to the conventional post-verification. By comparison, ``Adapt to Re-rank'' is directly applied to the coarse-grained search result; in other words, the re-ranking is virtually moved ahead to process a much larger group of candidates. Thus, it may better preserve the ads with high re-ranking scores and benefit the online performance. Note that ``Adapt to Re-rank'' also contributes to the efficiency, as the time cost can be saved from the conventional re-ranking step. 



\begin{table}[t]
    \centering
    \small
    \begin{tabular}{p{2.4cm} | C{2.0cm} C{2.0cm}  }
    \ChangeRT{1pt} 
    \textbf{Method} & 
    \textbf{NDCG} & \textbf{MRR} 
    \\ 
    \hline
    Base & 0.8568 & 0.4140 \\
    Base + Re-rank & 0.9033 & 0.4202 \\
    Adapt to Re-rank & \textbf{0.9096} & \textbf{0.4309} \\
    \ChangeRT{1pt}
    \end{tabular}
    \caption{Impact of re-rank oriented adaptation to the post re-rank retrieval quality.} 
    \vspace{-20pt}
    \label{tab:2}
\end{table}

\begin{table}[t]
    \centering
    \small
    \begin{tabular}{p{1.2cm} | C{1.8cm} C{1.8cm} C{1.8cm} }
    \ChangeRT{1pt} 
    \textbf{Method} & 
    \textbf{Recall@100} & \textbf{Recall@500} & \textbf{Recall@1000} 
    \\ 
    \hline
    PQ & 0.3167 & 0.5058 & 0.6000 \\
    OPQ & 0.3320 & 0.5172 & 0.6140 \\
    ScaNN & 0.4288 & 0.6396 & 0.7218 \\
    MoPQ & 0.5004 & 0.6963 & 0.7659 \\
    \ChangeRT{1pt}
    \end{tabular}
    \caption{First-stage retrieval result (coarse-grained result before post-verification) from different compression methods.}
    \vspace{-10pt}
    \label{tab:3}
\end{table} 

\subsubsection{Impact from MoPQ} 
The comparison between MoPQ and other commonly used PQ compression methods: PQ (the default option in DiskANN), OPQ, and ScaNN, is shown as Table \ref{tab:3}. The size of codebooks is 64 bit (8 codebooks in total, each one has 256 codewords). We may observe that MoPQ outperforms the baselines with notable advantages, verifying that the first-stage retrieval quality can be substantially improved from it. 

\subsubsection{Online A/B Test} The online A/B tests about Uni-Retriever are reported in Table \ref{tab:4}. Since our techniques are progressively developed, there are three versions shipped to Bing's production step-by-step. Particularly, the V1 version is highlighted for being ``Multi-Objective'': the contrastive learning is combined with knowledge distillation to optimize both CTR and relevance; meanwhile, the embeddings are disentangled for both objectives. This version achieves +3.27\% RPM growth in A/B test, with a P-Value of 1.82$e-2$ (P-Value below 5$e-2$ is considered as significant in the production). The V2 version improves the negative sampling strategy on top of the V1 version (with cross-device negative sampling, and ANN hard negatives filtered by relevance). Such an improvement gives rise to +3.06\% RPM growth (with V1 version already mainstreamed in production). Finally, with the EBR serving pipeline improved by the optimized DiskANN, the V3 version further leads to another +1.63\% RPM gain after the mainstream of the previous two versions. 

\begin{table}[t]
    \centering
    \small
    \begin{tabular}{p{3.0cm} | C{2.0cm} C{2.0cm} }
    \ChangeRT{1pt} 
    \textbf{Shipped Version} & 
    \textbf{RPM Growth} & \textbf{P-Value}
    \\ 
    \hline
    V1 (Multi-Objective)   & +3.27\% & 1.82$e^{-2}$ \\
    V2 (Negative Sampling) & +3.06\%	& 1.57$e^{-2}$ \\ 
    V3 (Optimized DiskANN) & +1.63\% & 2.06$e^{-2}$ \\ 
    \ChangeRT{1pt}
    \end{tabular}
    \caption{Online A/B Tests for Uni-Retriever (V1-V3).}
    \vspace{-10pt}
    \label{tab:4}
\end{table} 

\section{Related Work}


EBR has been widely applied to many applications, e.g., web search \cite{shen2014learning,liu2021pre}, question answering \cite{karpukhin2020dense,qu2021rocketqa}, online advertising \cite{lu2020twinbert,li2021adsgnn}, and recommender systems \cite{huang2020embedding,xiao2021training}. Generic EBR systems are designed for homogeneous objectives; e.g., the items users' may click in recommender systems, or the web-pages containing ground-truth answers in web search. However, EBR in sponsored search is more complicated due to the need of serving multiple retrieval purposes. For one thing, the retrieved ads have to be semantically close to user's specified query. For another thing, the retrieved ads are expected to be clicked by the user with high probabilities. Such a problem was recognized by prior works \cite{fan2019mobius,lu2020twinbert,li2019learning}. However, the learning of a unified model for the retrieval of ads with both high-CTR and high relevance still lacks of systematic studies.  

A typical EBR system consists of two fundamental components: the representation model and the ANN index. In Recent years, the pretrained language models (PLMs) \cite{devlin2018bert} are widely adopted as the backbone of representation models \cite{reimers2019sentence,karpukhin2020dense,luan2020sparse}; besides, people also propose to learn specific PLMs for EBR purpose \cite{chang2020pre,gao2021condenser,lu2021less}. In addition to the backbone model, the training algorithm is another critical factor. Most of the representation model is trained via contrastive learning \cite{hadsell2006dimensionality}, whose performance is affected by the negative sampling strategy. On one hand, it is found that the representation quality can be improved by using an increased amount of negative samples \cite{chen2020simple,he2020momentum,qu2021rocketqa}. As a result, different methods are used to improve the scale of negative samples in a cost-efficient way, e.g., in-batch negative sampling \cite{chen2020simple,karpukhin2020dense}, and momentum encoder \cite{he2020momentum}. On the other hand, the representation quality may also benefit from hard negatives. It's found that the negative samples selected by lexical similarity \cite{luan2020sparse,karpukhin2020dense} are helpful; later on, people observed additional gains by using negatives from ANN \cite{gillick2019learning,xiong2020approximate}. 

The ANN index is another building block of EBR in addition to the representation model, with which the Top-$N$ answers can be efficiently retrieved from the entire corpus \cite{jegou2011searching,ge2013optimized,malkov2018efficient,fu2017fast}. However, the existing methods are limited by either time cost, memory usage, or retrieval quality when applied to massive-scale EBR. The latest studies, e.g., DiskANN \cite{subramanya2019diskann}, SPANN \cite{chen2021spann}, propose to jointly leverage embedding compression and graph-based index for time and memory scalable ANN. Knowing that Bing needs to work with a billion-scale ads corpus, we leverage DiskANN for its high scalability, and we move beyond by substantially optimizing the compressed embeddings and posting lists for better retrieval quality.

\section{Conclusion} 
In this paper, we introduce our representation learning framework, Uni-Retriever, developed for Bing Sponsored Search. In Uni-Retriever, the knowledge distillation and contrastive learning are collaborated for the retrieval of ads with both high relevance and high CTR. We elaborate our optimization strategies designed for Uni-Retriever, including the cross-device negative sampling, relevance filtered ANN hard negatives, and disentangled representation learning. We also present our practice on massive-scale EBR serving, which further contributes to the retrieval performance on top of the substantially optimized DiskANN. Finally, we perform extensive offline experiments and online A/B tests, whose results verify the effectiveness of the proposed techniques. 



\bibliographystyle{ACM-Reference-Format}
\bibliography{ref}


\begin{thebibliography}{39}


\ifx \showCODEN    \undefined \def \showCODEN     #1{\unskip}     \fi
\ifx \showDOI      \undefined \def \showDOI       #1{#1}\fi
\ifx \showISBNx    \undefined \def \showISBNx     #1{\unskip}     \fi
\ifx \showISBNxiii \undefined \def \showISBNxiii  #1{\unskip}     \fi
\ifx \showISSN     \undefined \def \showISSN      #1{\unskip}     \fi
\ifx \showLCCN     \undefined \def \showLCCN      #1{\unskip}     \fi
\ifx \shownote     \undefined \def \shownote      #1{#1}          \fi
\ifx \showarticletitle \undefined \def \showarticletitle #1{#1}   \fi
\ifx \showURL      \undefined \def \showURL       {\relax}        \fi
\providecommand\bibfield[2]{#2}
\providecommand\bibinfo[2]{#2}
\providecommand\natexlab[1]{#1}
\providecommand\showeprint[2][]{arXiv:#2}

\bibitem[Baranchuk et~al\mbox{.}(2018)]%
        {baranchuk2018revisiting}
\bibfield{author}{\bibinfo{person}{Dmitry Baranchuk}, \bibinfo{person}{Artem
  Babenko}, {and} \bibinfo{person}{Yury Malkov}.}
  \bibinfo{year}{2018}\natexlab{}.
\newblock \showarticletitle{Revisiting the inverted indices for billion-scale
  approximate nearest neighbors}. In \bibinfo{booktitle}{\emph{Proceedings of
  the European Conference on Computer Vision (ECCV)}}.
  \bibinfo{pages}{202--216}.
\newblock


\bibitem[Chang et~al\mbox{.}(2020)]%
        {chang2020pre}
\bibfield{author}{\bibinfo{person}{Wei-Cheng Chang}, \bibinfo{person}{Felix~X
  Yu}, \bibinfo{person}{Yin-Wen Chang}, \bibinfo{person}{Yiming Yang}, {and}
  \bibinfo{person}{Sanjiv Kumar}.} \bibinfo{year}{2020}\natexlab{}.
\newblock \showarticletitle{Pre-training tasks for embedding-based large-scale
  retrieval}.
\newblock \bibinfo{journal}{\emph{arXiv preprint arXiv:2002.03932}}
  (\bibinfo{year}{2020}).
\newblock


\bibitem[Chen et~al\mbox{.}(2021)]%
        {chen2021spann}
\bibfield{author}{\bibinfo{person}{Qi Chen}, \bibinfo{person}{Bing Zhao},
  \bibinfo{person}{Haidong Wang}, \bibinfo{person}{Mingqin Li},
  \bibinfo{person}{Chuanjie Liu}, \bibinfo{person}{Zengzhong Li},
  \bibinfo{person}{Mao Yang}, {and} \bibinfo{person}{Jingdong Wang}.}
  \bibinfo{year}{2021}\natexlab{}.
\newblock \showarticletitle{SPANN: Highly-efficient Billion-scale Approximate
  Nearest Neighbor Search}.
\newblock \bibinfo{journal}{\emph{arXiv preprint arXiv:2111.08566}}
  (\bibinfo{year}{2021}).
\newblock


\bibitem[Chen et~al\mbox{.}(2020)]%
        {chen2020simple}
\bibfield{author}{\bibinfo{person}{Ting Chen}, \bibinfo{person}{Simon
  Kornblith}, \bibinfo{person}{Mohammad Norouzi}, {and}
  \bibinfo{person}{Geoffrey Hinton}.} \bibinfo{year}{2020}\natexlab{}.
\newblock \showarticletitle{A simple framework for contrastive learning of
  visual representations}. In \bibinfo{booktitle}{\emph{International
  conference on machine learning}}. PMLR, \bibinfo{pages}{1597--1607}.
\newblock


\bibitem[Devlin et~al\mbox{.}(2018)]%
        {devlin2018bert}
\bibfield{author}{\bibinfo{person}{Jacob Devlin}, \bibinfo{person}{Ming-Wei
  Chang}, \bibinfo{person}{Kenton Lee}, {and} \bibinfo{person}{Kristina
  Toutanova}.} \bibinfo{year}{2018}\natexlab{}.
\newblock \showarticletitle{Bert: Pre-training of deep bidirectional
  transformers for language understanding}.
\newblock \bibinfo{journal}{\emph{arXiv preprint arXiv:1810.04805}}
  (\bibinfo{year}{2018}).
\newblock


\bibitem[Fan et~al\mbox{.}(2019)]%
        {fan2019mobius}
\bibfield{author}{\bibinfo{person}{Miao Fan}, \bibinfo{person}{Jiacheng Guo},
  \bibinfo{person}{Shuai Zhu}, \bibinfo{person}{Shuo Miao},
  \bibinfo{person}{Mingming Sun}, {and} \bibinfo{person}{Ping Li}.}
  \bibinfo{year}{2019}\natexlab{}.
\newblock \showarticletitle{MOBIUS: towards the next generation of query-ad
  matching in baidu's sponsored search}. In
  \bibinfo{booktitle}{\emph{Proceedings of the 25th ACM SIGKDD International
  Conference on Knowledge Discovery \& Data Mining}}.
  \bibinfo{pages}{2509--2517}.
\newblock


\bibitem[Fu et~al\mbox{.}(2017)]%
        {fu2017fast}
\bibfield{author}{\bibinfo{person}{Cong Fu}, \bibinfo{person}{Chao Xiang},
  \bibinfo{person}{Changxu Wang}, {and} \bibinfo{person}{Deng Cai}.}
  \bibinfo{year}{2017}\natexlab{}.
\newblock \showarticletitle{Fast approximate nearest neighbor search with the
  navigating spreading-out graph}.
\newblock \bibinfo{journal}{\emph{arXiv preprint arXiv:1707.00143}}
  (\bibinfo{year}{2017}).
\newblock


\bibitem[Gao and Callan(2021)]%
        {gao2021condenser}
\bibfield{author}{\bibinfo{person}{Luyu Gao} {and} \bibinfo{person}{Jamie
  Callan}.} \bibinfo{year}{2021}\natexlab{}.
\newblock \showarticletitle{Condenser: a Pre-training Architecture for Dense
  Retrieval}. In \bibinfo{booktitle}{\emph{EMNLP}}. \bibinfo{pages}{981--993}.
\newblock


\bibitem[Ge et~al\mbox{.}(2013)]%
        {ge2013optimized}
\bibfield{author}{\bibinfo{person}{Tiezheng Ge}, \bibinfo{person}{Kaiming He},
  \bibinfo{person}{Qifa Ke}, {and} \bibinfo{person}{Jian Sun}.}
  \bibinfo{year}{2013}\natexlab{}.
\newblock \showarticletitle{Optimized product quantization}.
\newblock \bibinfo{journal}{\emph{IEEE transactions on pattern analysis and
  machine intelligence}} \bibinfo{volume}{36}, \bibinfo{number}{4}
  (\bibinfo{year}{2013}), \bibinfo{pages}{744--755}.
\newblock


\bibitem[Gillick et~al\mbox{.}(2019)]%
        {gillick2019learning}
\bibfield{author}{\bibinfo{person}{Daniel Gillick}, \bibinfo{person}{Sayali
  Kulkarni}, \bibinfo{person}{Larry Lansing}, \bibinfo{person}{Alessandro
  Presta}, \bibinfo{person}{Jason Baldridge}, \bibinfo{person}{Eugene Ie},
  {and} \bibinfo{person}{Diego Garcia-Olano}.} \bibinfo{year}{2019}\natexlab{}.
\newblock \showarticletitle{Learning dense representations for entity
  retrieval}.
\newblock \bibinfo{journal}{\emph{arXiv preprint arXiv:1909.10506}}
  (\bibinfo{year}{2019}).
\newblock


\bibitem[Guo et~al\mbox{.}(2020)]%
        {guo2020accelerating}
\bibfield{author}{\bibinfo{person}{Ruiqi Guo}, \bibinfo{person}{Philip Sun},
  \bibinfo{person}{Erik Lindgren}, \bibinfo{person}{Quan Geng},
  \bibinfo{person}{David Simcha}, \bibinfo{person}{Felix Chern}, {and}
  \bibinfo{person}{Sanjiv Kumar}.} \bibinfo{year}{2020}\natexlab{}.
\newblock \showarticletitle{Accelerating large-scale inference with anisotropic
  vector quantization}. In \bibinfo{booktitle}{\emph{ICML}}. PMLR,
  \bibinfo{pages}{3887--3896}.
\newblock


\bibitem[Hadsell et~al\mbox{.}(2006)]%
        {hadsell2006dimensionality}
\bibfield{author}{\bibinfo{person}{Raia Hadsell}, \bibinfo{person}{Sumit
  Chopra}, {and} \bibinfo{person}{Yann LeCun}.}
  \bibinfo{year}{2006}\natexlab{}.
\newblock \showarticletitle{Dimensionality reduction by learning an invariant
  mapping}. In \bibinfo{booktitle}{\emph{2006 IEEE Computer Society Conference
  on Computer Vision and Pattern Recognition (CVPR'06)}},
  Vol.~\bibinfo{volume}{2}. IEEE, \bibinfo{pages}{1735--1742}.
\newblock


\bibitem[He et~al\mbox{.}(2020)]%
        {he2020momentum}
\bibfield{author}{\bibinfo{person}{Kaiming He}, \bibinfo{person}{Haoqi Fan},
  \bibinfo{person}{Yuxin Wu}, \bibinfo{person}{Saining Xie}, {and}
  \bibinfo{person}{Ross Girshick}.} \bibinfo{year}{2020}\natexlab{}.
\newblock \showarticletitle{Momentum contrast for unsupervised visual
  representation learning}. In \bibinfo{booktitle}{\emph{Proceedings of the
  IEEE/CVF Conference on Computer Vision and Pattern Recognition}}.
  \bibinfo{pages}{9729--9738}.
\newblock


\bibitem[Hinton et~al\mbox{.}(2015)]%
        {hinton2015distilling}
\bibfield{author}{\bibinfo{person}{Geoffrey Hinton}, \bibinfo{person}{Oriol
  Vinyals}, {and} \bibinfo{person}{Jeff Dean}.}
  \bibinfo{year}{2015}\natexlab{}.
\newblock \showarticletitle{Distilling the knowledge in a neural network}.
\newblock \bibinfo{journal}{\emph{arXiv preprint arXiv:1503.02531}}
  (\bibinfo{year}{2015}).
\newblock


\bibitem[Huang et~al\mbox{.}(2020)]%
        {huang2020embedding}
\bibfield{author}{\bibinfo{person}{Jui-Ting Huang}, \bibinfo{person}{Ashish
  Sharma}, \bibinfo{person}{Shuying Sun}, \bibinfo{person}{Li Xia},
  \bibinfo{person}{David Zhang}, \bibinfo{person}{Philip Pronin},
  \bibinfo{person}{Janani Padmanabhan}, \bibinfo{person}{Giuseppe Ottaviano},
  {and} \bibinfo{person}{Linjun Yang}.} \bibinfo{year}{2020}\natexlab{}.
\newblock \showarticletitle{Embedding-based retrieval in facebook search}. In
  \bibinfo{booktitle}{\emph{Proceedings of the 26th ACM SIGKDD International
  Conference on Knowledge Discovery \& Data Mining}}.
  \bibinfo{pages}{2553--2561}.
\newblock


\bibitem[Jegou et~al\mbox{.}(2010)]%
        {jegou2010product}
\bibfield{author}{\bibinfo{person}{Herve Jegou}, \bibinfo{person}{Matthijs
  Douze}, {and} \bibinfo{person}{Cordelia Schmid}.}
  \bibinfo{year}{2010}\natexlab{}.
\newblock \showarticletitle{Product quantization for nearest neighbor search}.
\newblock \bibinfo{journal}{\emph{IEEE transactions on pattern analysis and
  machine intelligence}} \bibinfo{volume}{33}, \bibinfo{number}{1}
  (\bibinfo{year}{2010}), \bibinfo{pages}{117--128}.
\newblock


\bibitem[J{\'e}gou et~al\mbox{.}(2011)]%
        {jegou2011searching}
\bibfield{author}{\bibinfo{person}{Herv{\'e} J{\'e}gou},
  \bibinfo{person}{Romain Tavenard}, \bibinfo{person}{Matthijs Douze}, {and}
  \bibinfo{person}{Laurent Amsaleg}.} \bibinfo{year}{2011}\natexlab{}.
\newblock \showarticletitle{Searching in one billion vectors: re-rank with
  source coding}. In \bibinfo{booktitle}{\emph{ICASSP}}. IEEE.
\newblock


\bibitem[Johnson et~al\mbox{.}(2019)]%
        {johnson2019billion}
\bibfield{author}{\bibinfo{person}{Jeff Johnson}, \bibinfo{person}{Matthijs
  Douze}, {and} \bibinfo{person}{Herv{\'e} J{\'e}gou}.}
  \bibinfo{year}{2019}\natexlab{}.
\newblock \showarticletitle{Billion-scale similarity search with gpus}.
\newblock \bibinfo{journal}{\emph{IEEE Transactions on Big Data}}
  (\bibinfo{year}{2019}).
\newblock


\bibitem[Karpukhin et~al\mbox{.}(2020)]%
        {karpukhin2020dense}
\bibfield{author}{\bibinfo{person}{Vladimir Karpukhin}, \bibinfo{person}{Barlas
  Oguz}, \bibinfo{person}{Sewon Min}, \bibinfo{person}{Patrick Lewis},
  \bibinfo{person}{Ledell Wu}, \bibinfo{person}{Sergey Edunov},
  \bibinfo{person}{Danqi Chen}, {and} \bibinfo{person}{Wen-tau Yih}.}
  \bibinfo{year}{2020}\natexlab{}.
\newblock \showarticletitle{Dense Passage Retrieval for Open-Domain Question
  Answering}. In \bibinfo{booktitle}{\emph{EMNLP}}.
  \bibinfo{pages}{6769--6781}.
\newblock


\bibitem[Kim(2014)]%
        {kim-2014-convolutional}
\bibfield{author}{\bibinfo{person}{Yoon Kim}.} \bibinfo{year}{2014}\natexlab{}.
\newblock \showarticletitle{Convolutional Neural Networks for Sentence
  Classification}. In \bibinfo{booktitle}{\emph{Proceedings of the 2014
  Conference on Empirical Methods in Natural Language Processing ({EMNLP})}}.
  \bibinfo{publisher}{Association for Computational Linguistics},
  \bibinfo{address}{Doha, Qatar}, \bibinfo{pages}{1746--1751}.
\newblock
\urldef\tempurl%
\url{https://doi.org/10.3115/v1/D14-1181}
\showDOI{\tempurl}


\bibitem[Li et~al\mbox{.}(2021)]%
        {li2021adsgnn}
\bibfield{author}{\bibinfo{person}{Chaozhuo Li}, \bibinfo{person}{Bochen Pang},
  \bibinfo{person}{Yuming Liu}, \bibinfo{person}{Hao Sun},
  \bibinfo{person}{Zheng Liu}, \bibinfo{person}{Xing Xie},
  \bibinfo{person}{Tianqi Yang}, \bibinfo{person}{Yanling Cui},
  \bibinfo{person}{Liangjie Zhang}, {and} \bibinfo{person}{Qi Zhang}.}
  \bibinfo{year}{2021}\natexlab{}.
\newblock \showarticletitle{AdsGNN: Behavior-Graph Augmented Relevance Modeling
  in Sponsored Search}.
\newblock \bibinfo{journal}{\emph{arXiv preprint arXiv:2104.12080}}
  (\bibinfo{year}{2021}).
\newblock


\bibitem[Li et~al\mbox{.}(2019)]%
        {li2019learning}
\bibfield{author}{\bibinfo{person}{Xue Li}, \bibinfo{person}{Zhipeng Luo},
  \bibinfo{person}{Hao Sun}, \bibinfo{person}{Jianjin Zhang},
  \bibinfo{person}{Weihao Han}, \bibinfo{person}{Xianqi Chu},
  \bibinfo{person}{Liangjie Zhang}, {and} \bibinfo{person}{Qi Zhang}.}
  \bibinfo{year}{2019}\natexlab{}.
\newblock \showarticletitle{Learning Fast Matching Models from Weak
  Annotations}. In \bibinfo{booktitle}{\emph{The World Wide Web Conference}}.
  \bibinfo{pages}{2985--2991}.
\newblock


\bibitem[Liu et~al\mbox{.}(2021a)]%
        {liu2021pre}
\bibfield{author}{\bibinfo{person}{Yiding Liu}, \bibinfo{person}{Weixue Lu},
  \bibinfo{person}{Suqi Cheng}, \bibinfo{person}{Daiting Shi},
  \bibinfo{person}{Shuaiqiang Wang}, \bibinfo{person}{Zhicong Cheng}, {and}
  \bibinfo{person}{Dawei Yin}.} \bibinfo{year}{2021}\natexlab{a}.
\newblock \showarticletitle{Pre-trained Language Model for Web-scale Retrieval
  in Baidu Search}. In \bibinfo{booktitle}{\emph{Proceedings of the 27th ACM
  SIGKDD Conference on Knowledge Discovery \& Data Mining}}.
  \bibinfo{pages}{3365--3375}.
\newblock


\bibitem[Liu et~al\mbox{.}(2021b)]%
        {liu2021que2search}
\bibfield{author}{\bibinfo{person}{Yiqun Liu}, \bibinfo{person}{Kaushik
  Rangadurai}, \bibinfo{person}{Yunzhong He}, \bibinfo{person}{Siddarth
  Malreddy}, \bibinfo{person}{Xunlong Gui}, \bibinfo{person}{Xiaoyi Liu}, {and}
  \bibinfo{person}{Fedor Borisyuk}.} \bibinfo{year}{2021}\natexlab{b}.
\newblock \showarticletitle{Que2search: Fast and accurate query and document
  understanding for search at facebook}. In
  \bibinfo{booktitle}{\emph{Proceedings of the 27th ACM SIGKDD Conference on
  Knowledge Discovery \& Data Mining}}. \bibinfo{pages}{3376--3384}.
\newblock


\bibitem[Lu et~al\mbox{.}(2021)]%
        {lu2021less}
\bibfield{author}{\bibinfo{person}{Shuqi Lu}, \bibinfo{person}{Di He},
  \bibinfo{person}{Chenyan Xiong}, \bibinfo{person}{Guolin Ke},
  \bibinfo{person}{Waleed Malik}, \bibinfo{person}{Zhicheng Dou},
  \bibinfo{person}{Paul Bennett}, \bibinfo{person}{Tieyan Liu}, {and}
  \bibinfo{person}{Arnold Overwijk}.} \bibinfo{year}{2021}\natexlab{}.
\newblock \showarticletitle{Less is More: Pre-train a Strong Text Encoder for
  Dense Retrieval Using a Weak Decoder}.
\newblock \bibinfo{journal}{\emph{arXiv preprint arXiv:2102.09206}}
  (\bibinfo{year}{2021}).
\newblock


\bibitem[Lu et~al\mbox{.}(2020)]%
        {lu2020twinbert}
\bibfield{author}{\bibinfo{person}{Wenhao Lu}, \bibinfo{person}{Jian Jiao},
  {and} \bibinfo{person}{Ruofei Zhang}.} \bibinfo{year}{2020}\natexlab{}.
\newblock \showarticletitle{Twinbert: Distilling knowledge to twin-structured
  compressed BERT models for large-scale retrieval}. In
  \bibinfo{booktitle}{\emph{Proceedings of the 29th ACM International
  Conference on Information \& Knowledge Management}}.
  \bibinfo{pages}{2645--2652}.
\newblock


\bibitem[Luan et~al\mbox{.}(2020)]%
        {luan2020sparse}
\bibfield{author}{\bibinfo{person}{Yi Luan}, \bibinfo{person}{Jacob
  Eisenstein}, \bibinfo{person}{Kristina Toutanova}, {and}
  \bibinfo{person}{Michael Collins}.} \bibinfo{year}{2020}\natexlab{}.
\newblock \showarticletitle{Sparse, dense, and attentional representations for
  text retrieval}.
\newblock \bibinfo{journal}{\emph{arXiv preprint arXiv:2005.00181}}
  (\bibinfo{year}{2020}).
\newblock


\bibitem[Luan et~al\mbox{.}(2021)]%
        {luan2021sparse}
\bibfield{author}{\bibinfo{person}{Yi Luan}, \bibinfo{person}{Jacob
  Eisenstein}, \bibinfo{person}{Kristina Toutanova}, {and}
  \bibinfo{person}{Michael Collins}.} \bibinfo{year}{2021}\natexlab{}.
\newblock \showarticletitle{Sparse, Dense, and Attentional Representations for
  Text Retrieval}.
\newblock \bibinfo{journal}{\emph{Transactions of the Association for
  Computational Linguistics}}  \bibinfo{volume}{9} (\bibinfo{year}{2021}),
  \bibinfo{pages}{329--345}.
\newblock


\bibitem[Malkov and Yashunin(2018)]%
        {malkov2018efficient}
\bibfield{author}{\bibinfo{person}{Yu~A Malkov} {and} \bibinfo{person}{Dmitry~A
  Yashunin}.} \bibinfo{year}{2018}\natexlab{}.
\newblock \showarticletitle{Efficient and robust approximate nearest neighbor
  search using hierarchical navigable small world graphs}.
\newblock \bibinfo{journal}{\emph{IEEE transactions on pattern analysis and
  machine intelligence}} \bibinfo{volume}{42}, \bibinfo{number}{4}
  (\bibinfo{year}{2018}), \bibinfo{pages}{824--836}.
\newblock


\bibitem[Qu et~al\mbox{.}(2021)]%
        {qu2021rocketqa}
\bibfield{author}{\bibinfo{person}{Yingqi Qu}, \bibinfo{person}{Yuchen Ding},
  \bibinfo{person}{Jing Liu}, \bibinfo{person}{Kai Liu},
  \bibinfo{person}{Ruiyang Ren}, \bibinfo{person}{Wayne~Xin Zhao},
  \bibinfo{person}{Daxiang Dong}, \bibinfo{person}{Hua Wu}, {and}
  \bibinfo{person}{Haifeng Wang}.} \bibinfo{year}{2021}\natexlab{}.
\newblock \showarticletitle{RocketQA: An optimized training approach to dense
  passage retrieval for open-domain question answering}. In
  \bibinfo{booktitle}{\emph{Proceedings of the 2021 Conference of the North
  American Chapter of the Association for Computational Linguistics: Human
  Language Technologies}}. \bibinfo{pages}{5835--5847}.
\newblock


\bibitem[Reimers and Gurevych(2019)]%
        {reimers2019sentence}
\bibfield{author}{\bibinfo{person}{Nils Reimers} {and} \bibinfo{person}{Iryna
  Gurevych}.} \bibinfo{year}{2019}\natexlab{}.
\newblock \showarticletitle{Sentence-bert: Sentence embeddings using siamese
  bert-networks}.
\newblock \bibinfo{journal}{\emph{arXiv preprint arXiv:1908.10084}}
  (\bibinfo{year}{2019}).
\newblock


\bibitem[Shen et~al\mbox{.}(2014)]%
        {shen2014learning}
\bibfield{author}{\bibinfo{person}{Yelong Shen}, \bibinfo{person}{Xiaodong He},
  \bibinfo{person}{Jianfeng Gao}, \bibinfo{person}{Li Deng}, {and}
  \bibinfo{person}{Gr{\'e}goire Mesnil}.} \bibinfo{year}{2014}\natexlab{}.
\newblock \showarticletitle{Learning semantic representations using
  convolutional neural networks for web search}. In
  \bibinfo{booktitle}{\emph{Proceedings of the 23rd international conference on
  world wide web}}.
\newblock


\bibitem[Subramanya et~al\mbox{.}(2019)]%
        {subramanya2019diskann}
\bibfield{author}{\bibinfo{person}{Suhas~Jayaram Subramanya},
  \bibinfo{person}{Rohan Kadekodi}, \bibinfo{person}{Ravishankar Krishaswamy},
  {and} \bibinfo{person}{Harsha~Vardhan Simhadri}.}
  \bibinfo{year}{2019}\natexlab{}.
\newblock \showarticletitle{Diskann: Fast accurate billion-point nearest
  neighbor search on a single node}. In \bibinfo{booktitle}{\emph{Proceedings
  of the 33rd International Conference on Neural Information Processing
  Systems}}. \bibinfo{pages}{13766--13776}.
\newblock


\bibitem[Wu et~al\mbox{.}(2016)]%
        {wu2016google}
\bibfield{author}{\bibinfo{person}{Yonghui Wu}, \bibinfo{person}{Mike
  Schuster}, \bibinfo{person}{Zhifeng Chen}, \bibinfo{person}{Quoc~V Le},
  \bibinfo{person}{Mohammad Norouzi}, \bibinfo{person}{Wolfgang Macherey},
  \bibinfo{person}{Maxim Krikun}, \bibinfo{person}{Yuan Cao},
  \bibinfo{person}{Qin Gao}, \bibinfo{person}{Klaus Macherey}, {et~al\mbox{.}}}
  \bibinfo{year}{2016}\natexlab{}.
\newblock \showarticletitle{Google's neural machine translation system:
  Bridging the gap between human and machine translation}.
\newblock \bibinfo{journal}{\emph{arXiv preprint arXiv:1609.08144}}
  (\bibinfo{year}{2016}).
\newblock


\bibitem[Xiao et~al\mbox{.}(2021a)]%
        {xiao2021training}
\bibfield{author}{\bibinfo{person}{Shitao Xiao}, \bibinfo{person}{Zheng Liu},
  \bibinfo{person}{Yingxia Shao}, \bibinfo{person}{Tao Di}, {and}
  \bibinfo{person}{Xing Xie}.} \bibinfo{year}{2021}\natexlab{a}.
\newblock \showarticletitle{Training Microsoft News Recommenders with
  Pretrained Language Models in the Loop}.
\newblock \bibinfo{journal}{\emph{arXiv e-prints}} (\bibinfo{year}{2021}),
  \bibinfo{pages}{arXiv--2102}.
\newblock


\bibitem[Xiao et~al\mbox{.}(2021b)]%
        {xiao2021match}
\bibfield{author}{\bibinfo{person}{Shitao Xiao}, \bibinfo{person}{Zheng Liu},
  \bibinfo{person}{Yingxia Shao}, \bibinfo{person}{Defu Lian}, {and}
  \bibinfo{person}{Xing Xie}.} \bibinfo{year}{2021}\natexlab{b}.
\newblock \showarticletitle{Matching-oriented Product Quantization For Ad-hoc
  Retrieval}.
\newblock \bibinfo{journal}{\emph{arXiv preprint arXiv:2104.07858}}
  (\bibinfo{year}{2021}).
\newblock


\bibitem[Xiong et~al\mbox{.}(2020)]%
        {xiong2020approximate}
\bibfield{author}{\bibinfo{person}{Lee Xiong}, \bibinfo{person}{Chenyan Xiong},
  \bibinfo{person}{Ye Li}, \bibinfo{person}{Kwok-Fung Tang},
  \bibinfo{person}{Jialin Liu}, \bibinfo{person}{Paul~N Bennett},
  \bibinfo{person}{Junaid Ahmed}, {and} \bibinfo{person}{Arnold Overwijk}.}
  \bibinfo{year}{2020}\natexlab{}.
\newblock \showarticletitle{Approximate Nearest Neighbor Negative Contrastive
  Learning for Dense Text Retrieval}. In
  \bibinfo{booktitle}{\emph{International Conference on Learning
  Representations}}.
\newblock


\bibitem[Xiong et~al\mbox{.}(2021)]%
        {xiong2021approximate}
\bibfield{author}{\bibinfo{person}{Lee Xiong}, \bibinfo{person}{Chenyan Xiong},
  \bibinfo{person}{Ye Li}, \bibinfo{person}{Kwok-Fung Tang},
  \bibinfo{person}{Jialin Liu}, \bibinfo{person}{Paul~N. Bennett},
  \bibinfo{person}{Junaid Ahmed}, {and} \bibinfo{person}{Arnold Overwijk}.}
  \bibinfo{year}{2021}\natexlab{}.
\newblock \showarticletitle{Approximate Nearest Neighbor Negative Contrastive
  Learning for Dense Text Retrieval}. In \bibinfo{booktitle}{\emph{ICLR}}.
\newblock


\bibitem[Yao et~al\mbox{.}(2021)]%
        {yao2021adapt}
\bibfield{author}{\bibinfo{person}{Yunzhi Yao}, \bibinfo{person}{Shaohan
  Huang}, \bibinfo{person}{Wenhui Wang}, \bibinfo{person}{Li Dong}, {and}
  \bibinfo{person}{Furu Wei}.} \bibinfo{year}{2021}\natexlab{}.
\newblock \showarticletitle{Adapt-and-Distill: Developing Small, Fast and
  Effective Pretrained Language Models for Domains}.
\newblock \bibinfo{journal}{\emph{arXiv preprint arXiv:2106.13474}}
  (\bibinfo{year}{2021}).
\newblock


\end{thebibliography}

\end{document}